\documentclass{elsart}

 \usepackage{epsfig}

\usepackage{amssymb,amsmath}

\begin{document}

\begin{frontmatter}



\title{ The Kauffman model on Small-World Topology}
\author{ Carlos Handrey A. Ferraz } and
\ead{handrey@fisica.ufc.br}
\author[ufc]{Hans J. Herrmann \corauthref{cor2}}
\ead{hans@fisica.ufc.br}
\corauth[cor2]{Corresponding Author.}
\address[ufc]{Departament of Physics, Universidade Federal do Ceara-UFC
Campus do Pici, PO Box 6030, CEP 60.455-760, Fortaleza, CE, Brazil}


\begin{abstract}
 We  apply  Kauffman's automata on  small-world networks
to study the crossover between the short-range  and the infinite-range case. We perform  accurate calculations  on square lattices to obtain  both critical exponents and fractal dimensions.  Particularly,  we find  an increase of the damage propagation  and a decrease in the fractal dimensions when adding long-range connections.
\end{abstract}

\end{frontmatter}
\section{Introduction}
\hspace{0.5cm} The Kauffman model \cite{kauff}
or more generally random Boolean networks have been
studied in the past to describe genetic regulatory networks but are in fact
very general because they do not assume any particular function of the nodes
and can be applied on any topology. While in the late eighties much work has
been done on the damage spreading transition renewed interest in this
model has risen in the last years because of the issues of sychronisation\cite{harvey}, stability  \cite{bilke}, 
control of chaos \cite{luque}  and investigations on scale-free topologies\cite {aldana,iguchi}.
A topology that is very characteristic for human networks
is the small-world as introduced by Watts and Strogatz \cite{watt}. It is the purpose
of this paper to present some numerical studies of the phase transition
on the square lattice, complementing work by Stauffer \cite{stauf}, and then on its
small-world variant showing the appearance of a new universality class. 

\hspace{0.5cm} Let us first  review  Kauffman cellular automata.
The Kauffman model   is a random mixture of all possible Booleaan rules.
Each of $N$ lattice sites hosts a Boolean variable $\sigma_{i}$ ( spin up or down) which
is either zero or unity. The time evolution of this model is determined by $N$ functions $f_{i}$
(rules) which are randomly chosen for each site independently, and by the choice of $K$ input
sites \{$j_{K}(i)$\} for each site $i$. Thus the value $\sigma_{i}$ 
at site $i$  for time $t$+1 is given by:
\begin{equation}
\sigma_{i}(t+1)=f_{i}(\sigma_{j_{1}}(t), \ldots,\sigma_{j_{K}}(t))\qquad  (i=1,2,\ldots,N)
\end{equation} 
Each Boolean function $f_{i}$ is specified, once its value is given for each of the $2^{K}$ possible 
neighbour configurations. Here we are concerned with those cases where both the inputs  and  the 
chosen Boolean functions do not change with time (quenched model).
\begin{figure}[h]
\centering
\includegraphics*[scale=0.40]{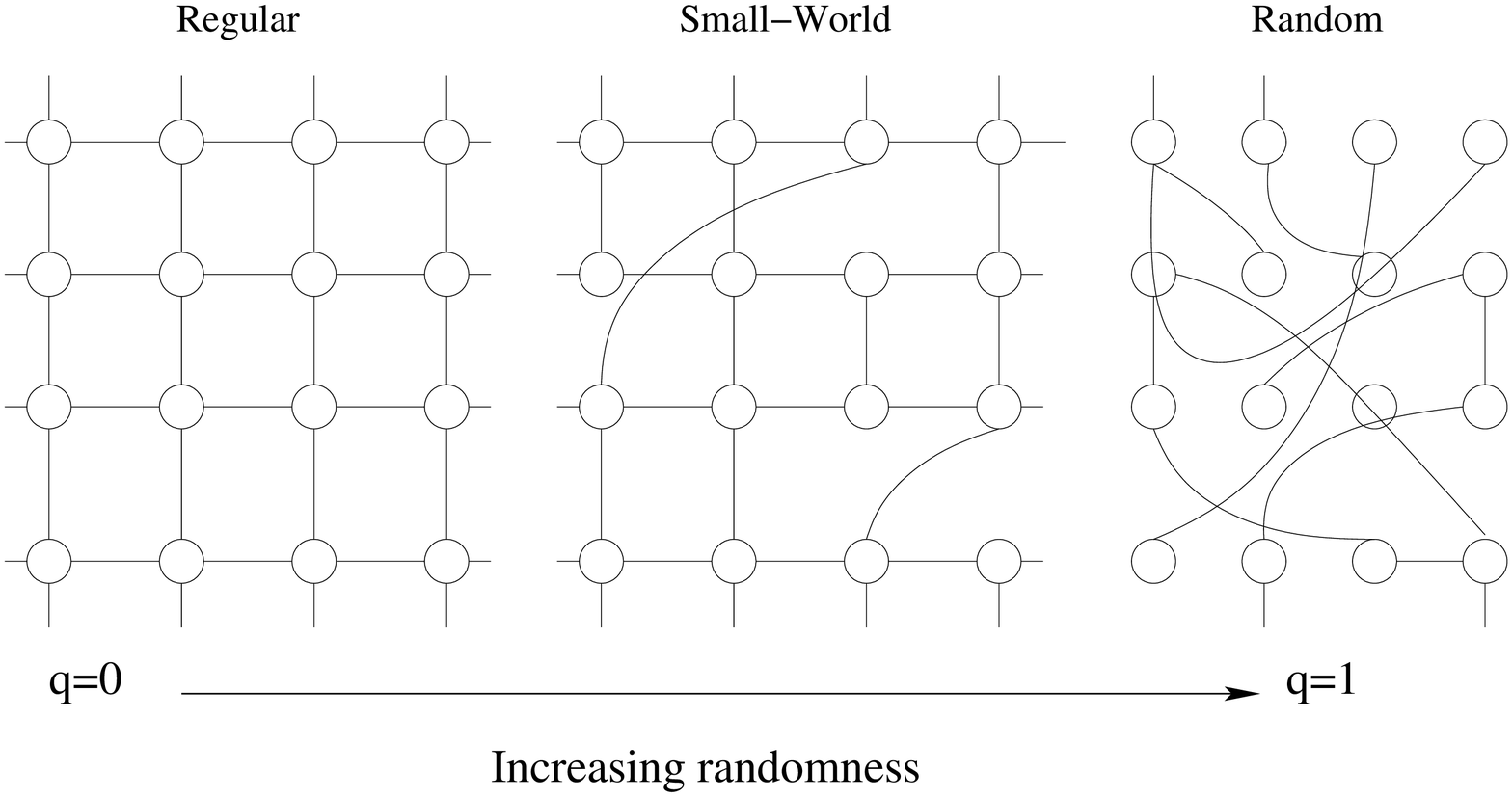}
\caption{ Rewiring procedure }
\label{figure 01}
\end{figure}

\hspace*{0.5cm}  The Kauffman model has usually been considered in the last years, on one hand with an infinite range of interaction, 
known as the `mean-field' approximation \cite{derr,drossel} in which for each site $i$ , the $K$ inputs are randomly chosen among the
$N$ spins. Such approximation was introduced to describe genetic interactions, to explain mutations and stability processes in 
biological genes.  Indeed, this was the original proposal of the Kauffman model.
On the other hand, this model has also been simulated on lattices with nearest-neighbour interactions, the so-called short-range
case \cite{stauf}. If one considers a square lattice, as we do here, for each site $i$ the $K$ inputs
are just the four nearest neighbours, thus there being $2^{4}=16$ neighbor configurations, and therefore 
$2^{16}=65536$ different possible rules. One can implement this  by introducing a 
probability $p$ of selecting a rule with the result spin up for each neighbor configuration and each  lattice site.  Of course, because of the symmetry of Boolean functions, one also has a probability 1-$p$ that  the result is spin down. \\
\hspace*{0.5 cm}In the past, the connection topology had been assumed  to be either completely random or completely regular. But many biological, technological and social networks lie somewhere between these two extremes. The  `small-world' topology seems to be very interesting for this purpose. In order to get more knowledge about this  intermediate range, we interpolate between  regular  and 
random networks by starting from a square lattice with $N$ sites and $K$=4 inputs of  nearest-neighbors and introducing a rewiring probability $q$ to each input $K$ connecting one site to any other, as indicated in Fig.1. Dynamical systems with small-world coupling exhibit enhanced signal-propagation speed and synchronizability. In particular, small 
damages, for example mutations on genetic material, spread more easily on small-world networks than on regular lattice. \\
\hspace*{0.5 cm} In  the present work, we have performed  simulations based on a parallel bit manipulation technique called {\it multi-spin 
coding} \cite{friedberg} to attempt to determine the critical point $p_c$ in analogy to percolation, both for the short-range case 
($q=0$) and for the small-world case ($q\neq0$).  To this end, we plot the normalized damage mass (see next section), i.e. our order paramenter $\psi$
, versus the parameter $p$.  Once determinated the critical points $p_c$ for each case,  we  evaluate the critical exponents and the fractal dimensions. \\

\section{`Damage Spreading' on the Square Lattice}
\begin{figure}[!h]
 \begin{minipage} [!l]{0.48\linewidth}
 \includegraphics*[angle=-90,width=\linewidth]{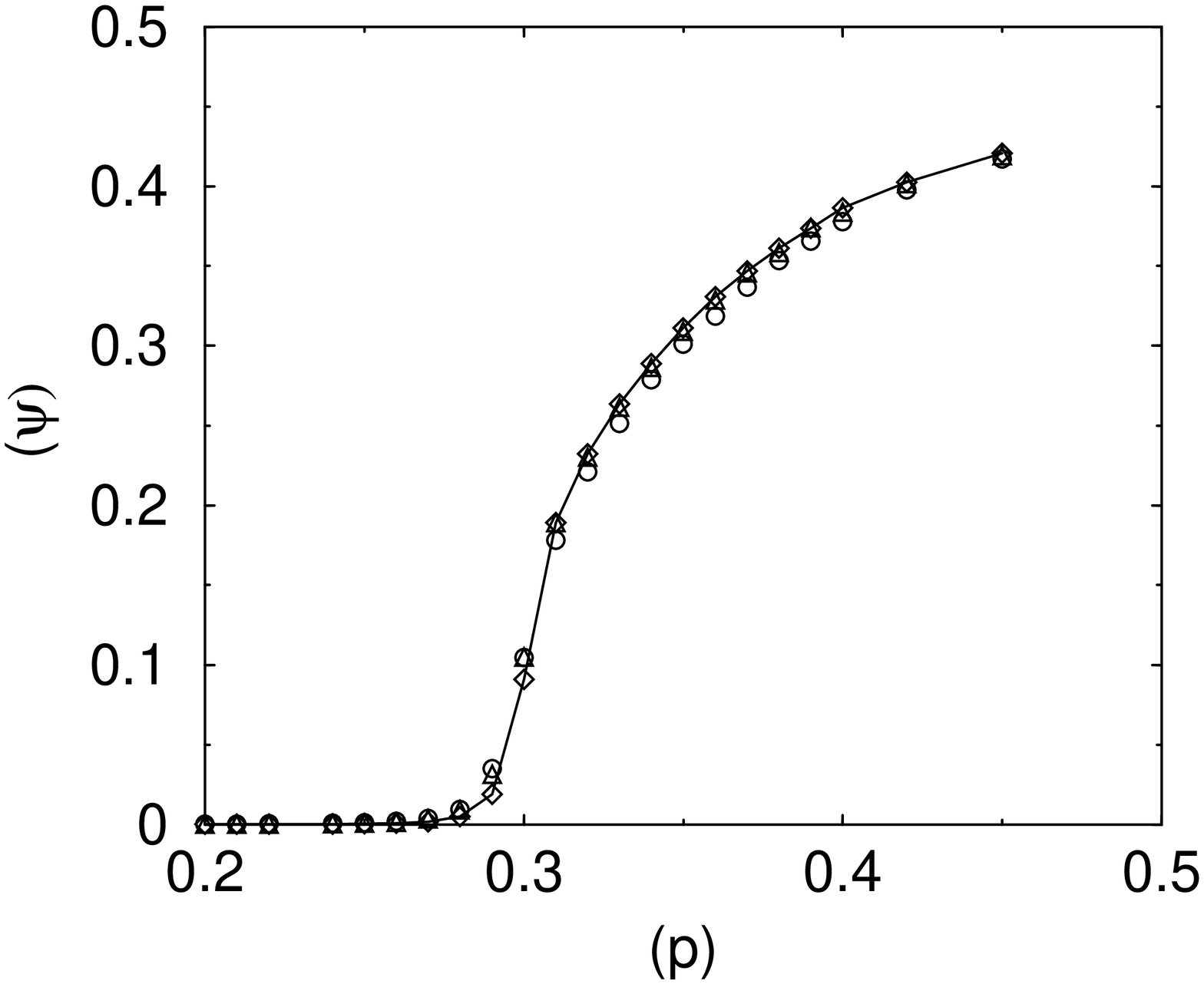}
 \caption{Plot of $\psi$ vs $p$ for the short-range case ($q=0.00$). Lattice sizes $L$=256; $\circ$, $L$=384; $\triangle$ and $L$=512; 
 $\Diamond$.}
 \end{minipage}\hfill
 \begin{minipage}[!l]{0.48\linewidth}
 \includegraphics*[angle=-90,width=\linewidth]{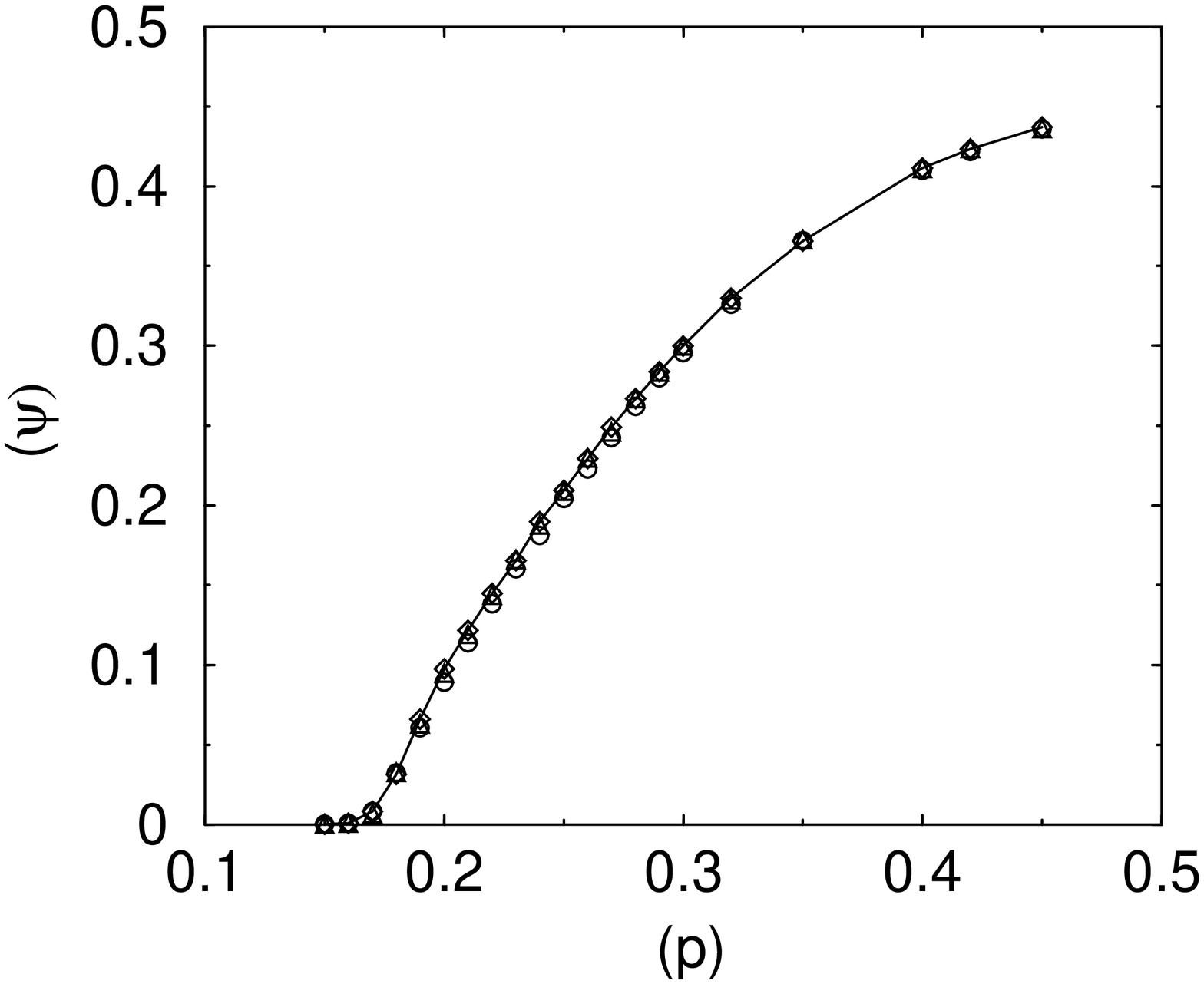}
 \caption{Plot of $\psi$ vs $p$ for the small-world case ($q=0.05$). Lattice sizes $L$=256; $\circ$, $L$=384; $\triangle$ and $L$=512; 
 $\Diamond$.}
 \end{minipage}
 \end{figure} 
\hspace*{0.5 cm}With the objective of understanding how a single mutation spreads through  genetic material, or how small hardware failures 
propagate on a computer architecture, we have compared two square lattices with identical sets of rules and almost  identical Boolean variables at 
each site. The only difference between them consists in a small modification of few central Boolean variables (typically less than 1\% 
of the number of sites) . So one can observe the time-development of the two lattices and check the difference between them after a long time  
for each  parameter value $ p$ . The difference between  lattices is measured as the number of `spins' that differ between the two lattices; this difference is also known as Hamming distance $d(t)$  between the lattice configurations \{$\sigma_{i}(t)$\} and  \{$\rho_{i}(t)$\}:
\begin{equation}
d(t)=\dfrac{1}{N}\sum_{i}|\sigma_{i}(t)-\rho_{i}(t)|.
\end{equation} 
 Following earlier works \cite{stauf,derr2,arcang}, we call this difference  the `damage'.  Further we define here the order 
 parameter of the system as:
 \begin{equation}
   \psi=\lim_{\substack{d(0)\rightarrow 0}}d(\infty),
  \end{equation}
 and one can also define a susceptibility as:
 \begin{equation}
 \chi= \dfrac{\partial d(\infty)} {\partial d(0)}.
 \label{ci}
 \end{equation}
  It has been observed that $\psi$ as well as  $ 1/ \chi$ go to zero  at 
 some critical concentration  $p_{c}$  in systems of dimensions greater than one for the short-range case Kauffman model similar to the
 para-ferromagnetic phase transition. In other words, for all $p \leqslant p_{c}$,  a small initial damage vanishes
 or remains small, i.e. belongs to a small cluster of `damaged spins',  after a sufficiently long time. One  says that the system is in the 
 frozen phase. On the other hand, for all $p>p_{c}$, a small initial damage spreads througout a 
 considerable part of the system. Then one says that the system is in the chaotic phase. Of particular interest is, however, the 
 border case $p=p_{c}$ where fractal properties appear  (see next section). To know more about phase 
 transitions in Kauffman's automata the reader should consult, for instance, Ref. \cite{derr2}. 
 \begin{figure}[!h]
 \centering
 \includegraphics*[scale=0.30,angle=-90]{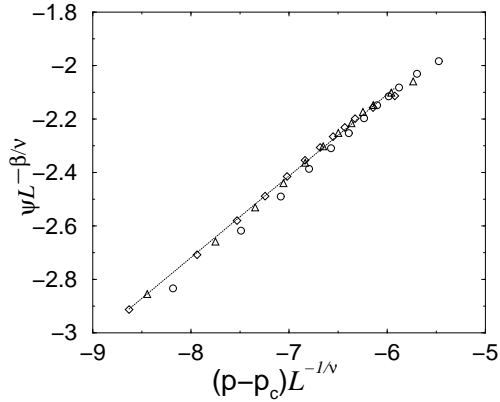}
 \caption{Log-log plot of $\psi L^{-\beta/ \nu}$ vs ($p-p_{c}$)$ L^{-1/ \nu}$ for the short-range case ($q=0.00$). 
 Lattice size $L$=256; $\circ$, $L$=384;$\triangle$ and $L$=512; $\Diamond$}
\end{figure}
  \begin{figure}[!h]
 \begin{minipage}[b]{0.48\linewidth}
 \includegraphics*[angle=-90,width=\linewidth]{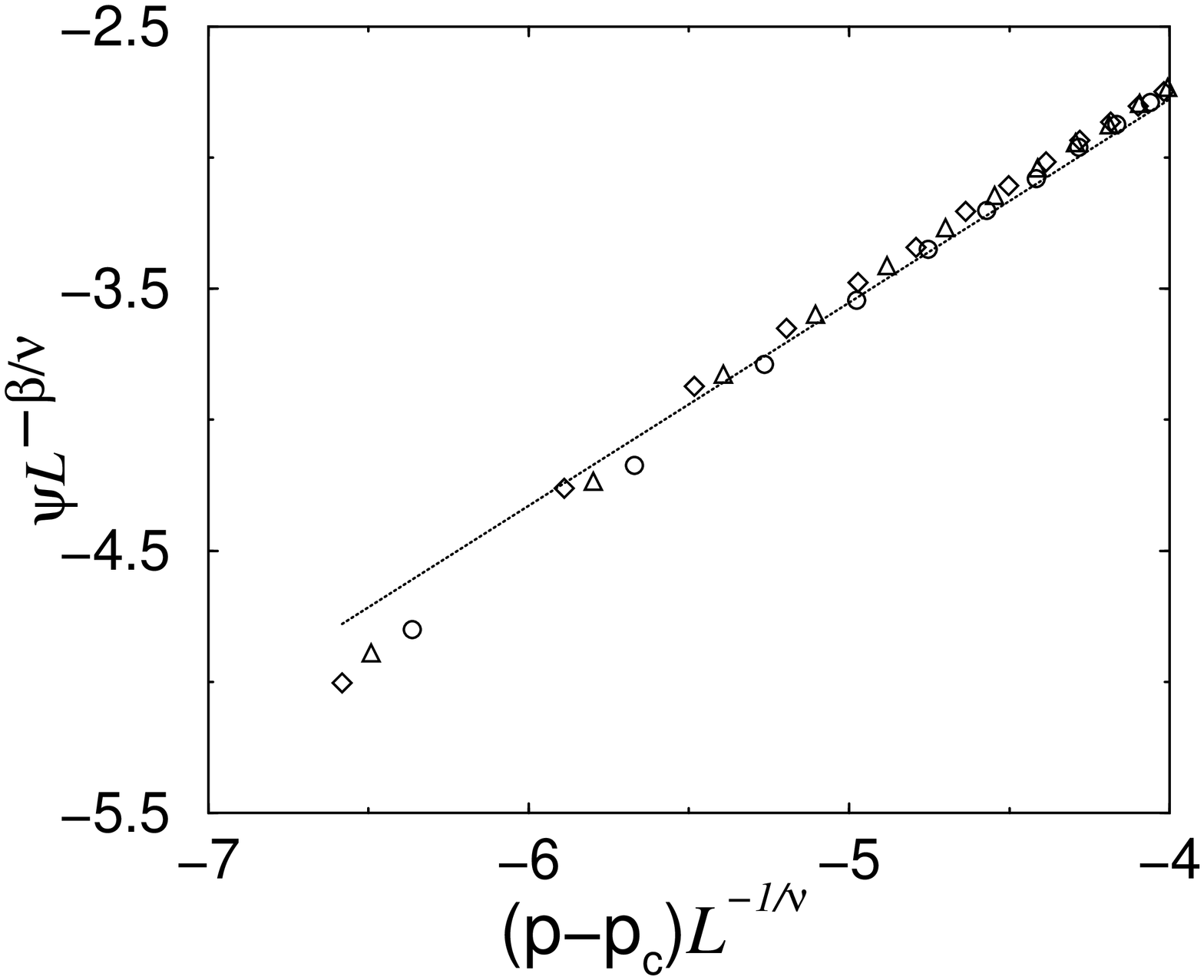}
 \caption{Log-log plot of $\psi L^{-\beta/ \nu}$ vs ($p-p_{c}$)$ L^{-1/ \nu}$ for the case ($q=0.05$). Lattice 
 size $L$=256; $\circ$, $L$=384;$\triangle$ and $L$=512; $\Diamond$}
 \end{minipage}\hfill
 \begin{minipage}[b]{0.48\linewidth}
 \includegraphics*[angle=-90,width=\linewidth]{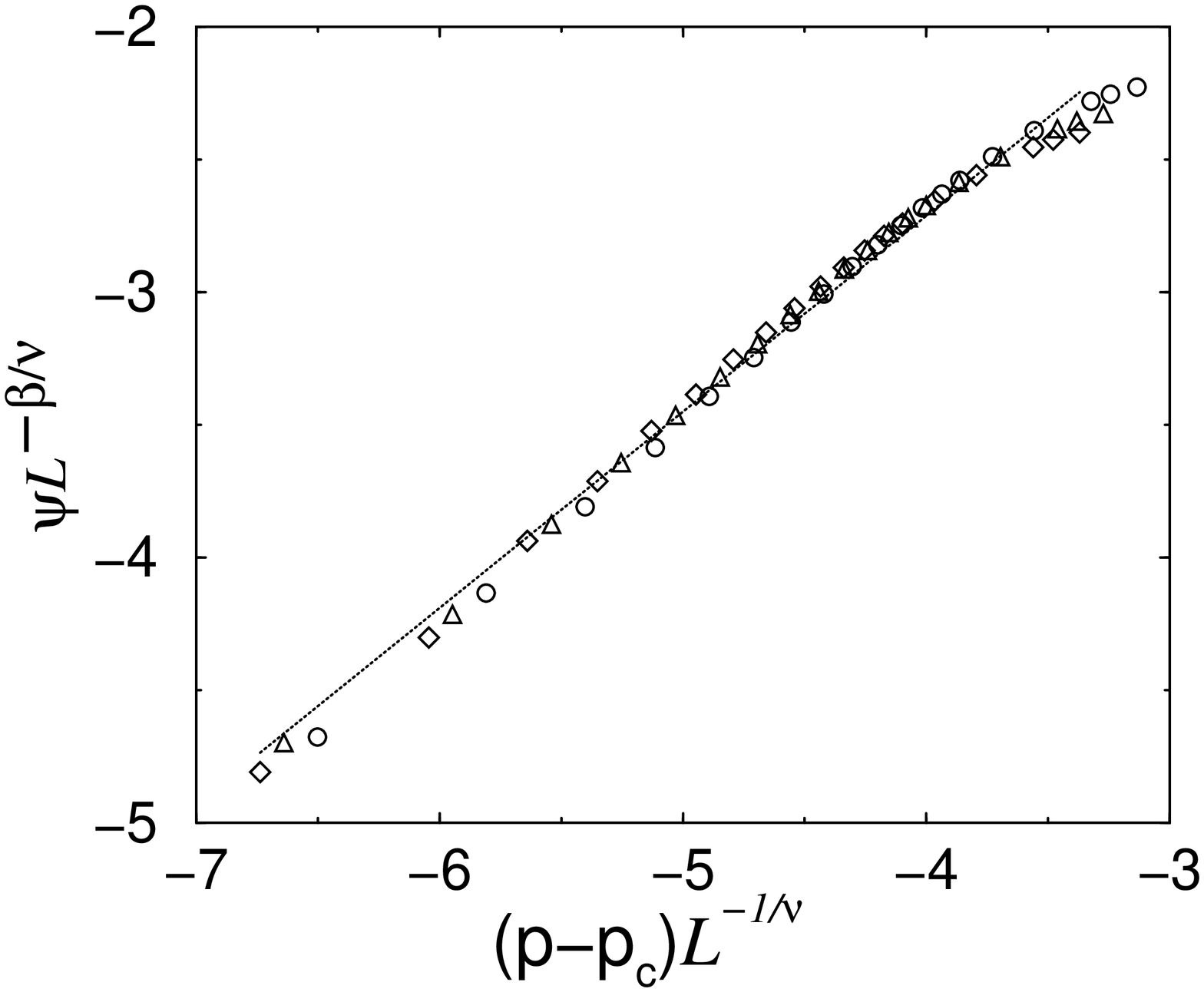}
 \caption{ Log-log plot of $\psi L^{-\beta/ \nu}$ vs ($p-p_{c}$)$ L^{-1/ \nu}$ for the  case ($q=0.10$). Lattice 
 size $L$=256; $\circ$, $L$=384;$\triangle$ and $L$=512; $\Diamond$}
 \end{minipage}
 \end{figure}
 \hspace*{0.5cm} As a way to find  $p_{c}$ in both the short-range and the small-world case, we plotted the order parameter 
 $\psi$ versus the parameter $p$ for the case $q=0 $(short-range) and $q=0.05$ and 0.10 (small-world) for three different lattice sizes 
 ($L$=256, 384 and 512) by flipping a small amount of  regularly spaced central spins  and taking an average over 100 runs for each 
 point. Everything  was calculated for up to $3\times10^3$ time-steps on an Intel Pentium IV processor. The updating speed using a {\it multi-spin coding} 
algorithm on an Intel machine which stores 64 spins in one computer word  was about 2.44 updates per nanosecond for two configurations following the same iteration rules. 
Our results are shown in Figs. 2 and 3, repectively for $q=0$ and  $q=0.05$. The transition between the frozen phase and the chaotic phase is seen to occur around $p_{c}=0.30 $ for the case $q=0$ and around $p_{c}=0.17$ and  $p_{c}=0.16$ for the cases $q=0.05$ and $q=0.10$, respectively. 
 \begin{center}
Table 1
\end{center}
\begin{table}[!h]
\begin{center}
\begin{tabular}{|c||c|c|c|} \hline
&$q=0.00$& $q=0.05$& $q=0.10$ \\ \hline \hline
$p_{c}$&0.30&0.17&0.16 \\ \hline
$\beta$ &$0.31\pm 0.04$&$0.78\pm 0.04$&$0.74\pm 0.04$\\ \hline
$\gamma$&$2.50\pm 0.15$&$4.75\pm0.20$&$4.65\pm0.15 $\\ \hline
\end{tabular}
\end{center}
\end{table}
  \hspace*{0.5cm}We  estimate  the critical exponents for each case by regarding  the fraction of  damaged sites ($\psi$) on each 
  side of the critical point after a sufficiently long time.  In the chaotic range ($p>p_{c}$), this fraction goes to zero as 
  ($p-p_{c})^\beta$. 
 While in the frozen range, the ratio of the final damage to the small initial damage ($\chi$) varies as ($p_{c}-p)^{-\gamma}$. We 
 have made a collapse of all data in order to check the consistency of the obtained results by making use of the following scaling 
 law: 
\begin{equation}
\psi(L,p)= L^{-\beta/ \nu}F((p-p_{c}) L^{-1/ \nu}),
\end{equation}
 where $L$ is the linear dimension of the lattice and $\nu$ is the exponent describing the divergence of the correlation length at  
 $p_{c}$. Fig. 4 shows the data collapse for the case $q=0$ at $p_{c}=0.30$ which is in good 
 agreement with Ref. \cite{stauf3}, whereas the Figs. 5 and 6 show the data collapse for the case $q=0.05 $ and $q=0.10$ at 
 $p_{c}=0.17$ and $p_{c}=0.16$ respectively.  Indeed, from Figs. 4, 5 and 6  we observe that  the points following the
 scaling relation given by Eq. 5 have a slope of  straight line equal to critical exponent  $\beta$ for each case.
 This could be an indication of the quality of our data. We have evaluated $\gamma$ 
 by means of the Eq. \ref{ci} at the transition threshold $p_{c}$. Our estimates together with the  $p_{c}$ values are shown  in  
 Table 1.


\section{Fractal Dimensions}
\hspace{0.5cm}In analogy with percolation theory, we have determined  the fractal dimensions d$_{act}$ and d$_{t}$ both for the short-range and the small-world cases by calculating the damage mass  as well as the touching time on the lattice bounderies for several lattice sizes $L$ at the critical point $p_{c}$. \\
The fractal dimension was defined by Stauffer \cite{stauf} through the asymptotic proportionalities ($L\rightarrow \infty$) in the following way:
\begin{equation}
<M>\sim L^{d_{act}}
\label{m}
\end{equation} 
\begin{equation}
<T_{t}>\sim L^{d_{t}},
\label{t}
\end{equation} 
where  Eq. \ref{m} says how the average damage increases  with the lattice size $L$  at the critical point, whereas Eq. \ref{t} says how the average time for the damage  spread  increases with lattice size $L$ at the critical point. The averages in Eqs. \ref{m} and \ref{t} are taken over those runs that eventually touch the  upper or  lower boundary of the square lattices. 
  \begin{figure}[!h]
 \begin{minipage}[b]{0.45\linewidth}
 \includegraphics*[angle=-90,width=\linewidth]{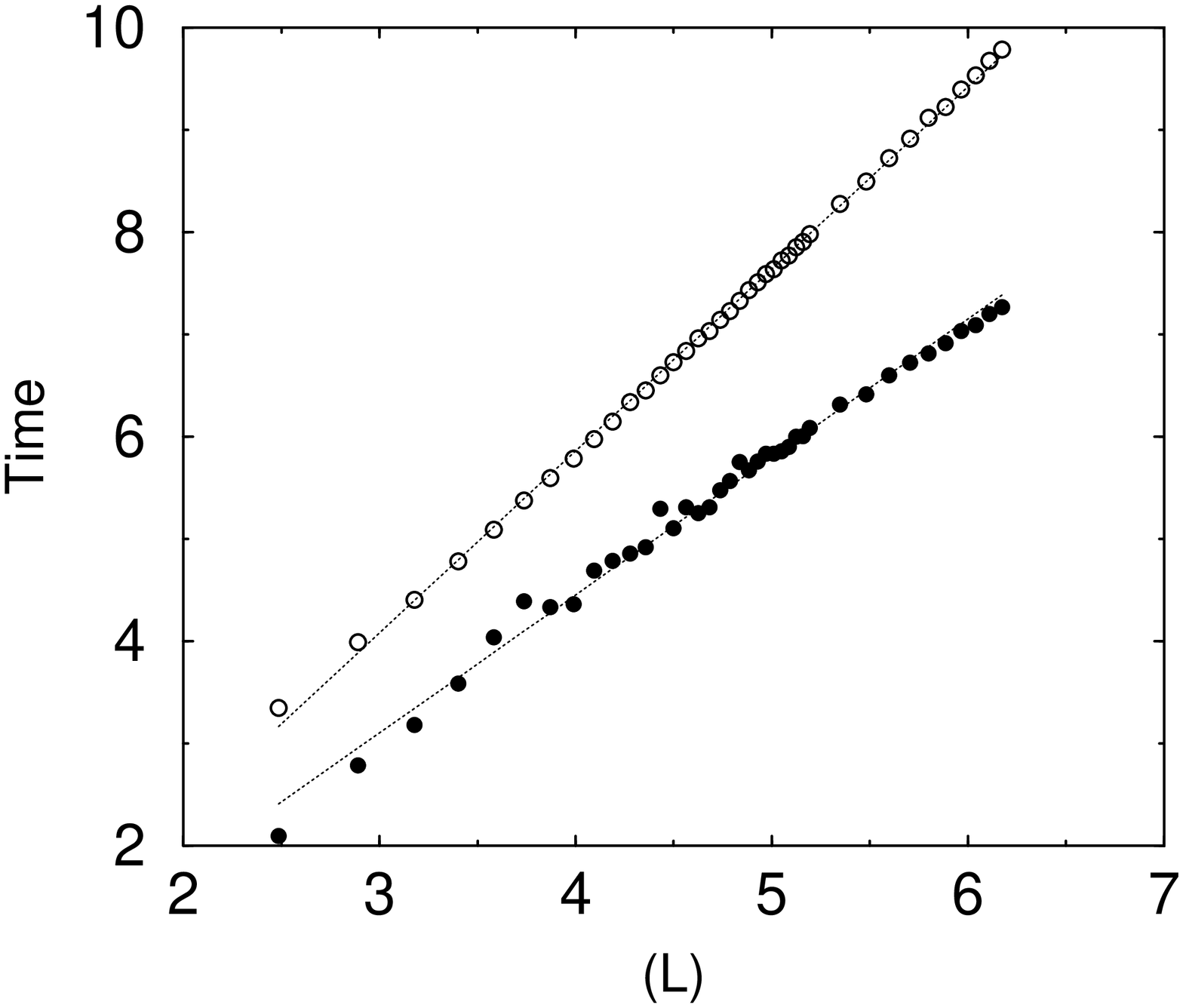}
 \caption{Log-log plot of  touching time vs $L$; $\bullet$, damage vs $L$; $\circ$ for the short-range case ($q=0.00$) }
 \end{minipage}\hfill
 \begin{minipage}[b]{0.45\linewidth}
 \includegraphics*[angle=-90,width=\linewidth]{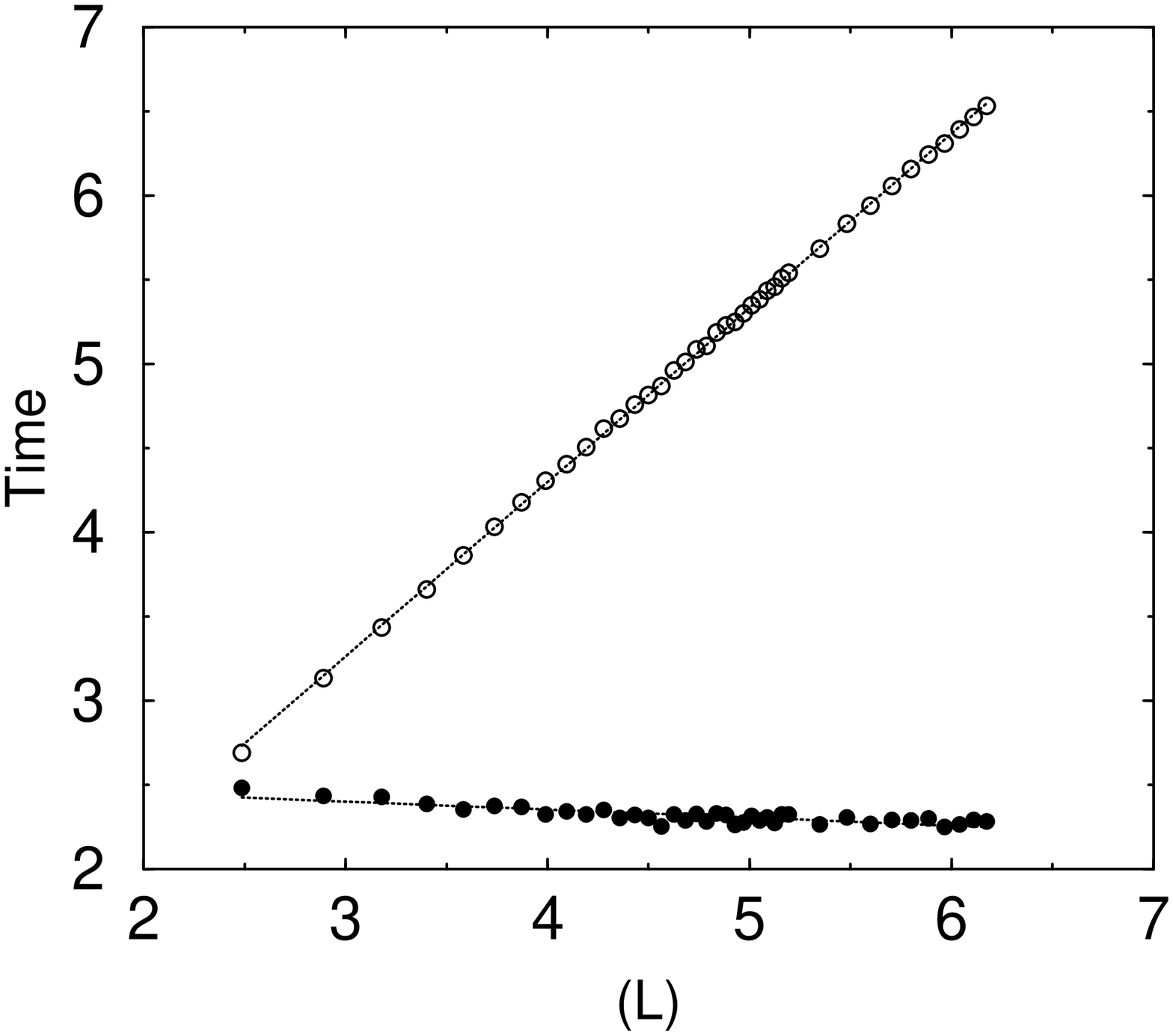}
 \caption{Log-log plot of  touching time vs $L$; $\bullet$, damage vs $L$; $\circ$ for the small-world case ($q=0.05$)}
 \end{minipage}
 \end{figure}
 \hspace*{0.5cm} In this paper as in Ref. \cite {corsten} we flipped a whole central line of spins at each time-step  and  let the 
 system  
 evolve until the final damage touches the lattice boundaries. Our results for the  touching time $T_{t}$ and the damage mass $M$ 
 at the  touching  time ( averaged  over 100 runs for $q=0$ and over 1000 runs for $q \neq 0$)  are shown as a log-log plot of  time 
 vs $L$  
 in Figs. 7 and 8, respectively for the  short-range ($q=0$) and the small-world ($q=0.05$) case. 
Our numerical results for the short-range case ($q=0$) turned out to be $d_{act}=1.79$ and $d_{t}=1.35$. These results are in good 
 agreement with Ref. \cite{stauf3} and obey the scaling relations: From table 1, we have $\beta=0.31$ and $\gamma=2.48$. 
 Assuming hyperscaling one has $d\nu=\gamma+2 \beta$, and we get $\nu=1.55$.  Standard percolation theory, suggests $ 
 d_{act}=d-\beta / \nu$, predicting $d_{act}=1.80$.  Our results for the  small-world case ($q=0.05$ and $q=0.10$) were $d_{act} 
 = 1.04$ and $d_{t}\approx 0$. Therefore one can see easily that at $q\neq0$ the system no longer obeys the scaling relations of  percolation.

\section{Conclusion}

\hspace{0.5cm}With the use of our algorithm based on the {\it multi-spin coding} technique, we simulate  Kauffman's automata, analyzing  how small central damages spread  on square lattices both in the short-range case and the small-world case. In this work, we have seen that for the small-world case the speed of damage spreading  is larger than for short-range case.  We  numerically evaluate critical exponents $\beta $ and $\gamma$ for  both cases.  We also estimate the fractal dimension as being  $d_{act}=1.79$  for the short-range case and $d_{act}=1.04 $  for small-world case.  Where for the latter case we can no longer observe  scaling relations for percolation. Finally, we expect that these and other more elaborated calculations will be helpful  to understand more general problems  concerning the  propagation of simple defects in complex systems.

\section{Acknowledgement}
 
 \hspace*{0.5cm} We thank CAPES, CNPq , FUNCAP and the Max Planck prize
for financial support.


\end{document}